\documentclass[11pt,letterpaper]{article}

\usepackage{epsfig}
\usepackage[margin=0.75in]{geometry}

\usepackage{amsmath}

\begin{document}




\title{Spin-dependent Bohm trajectories for hydrogen eigenstates}


\author{C. Colijn and E.R. Vrscay
\\Department of Applied Mathematics\\University of Waterloo
\\Waterloo, Ontario, Canada  N2L 3G1}
\maketitle



\begin{abstract}
The Bohm trajectories for several hydrogen atom eigenstates are determined,
taking into account the additional momentum term 
$\nabla \rho \times {\bf s}$ that arises from the Pauli current.
Unlike the original Bohmian result, the spin-dependent term yields
nonstationary trajectories. The
relationship between the trajectories and the standard
visualizations of orbitals is discussed.
The trajectories for a model problem
that simulates a $1s$-$2p$ transition in hydrogen are also examined.
\end{abstract}

%
%

\section{Introduction}
In David Bohm's original causal interpretation
of quantum mechanics \cite{Bo52}, the motion of a quantum
mechanical particle is determined by its wavefunction
$\psi$, which acts as a guidance wave \cite{deBr60}.
If the wavefunction is written as 
\begin{equation}
\label{bohmpsi}
\psi({\bf x},t)=R({\bf x},t)e^{iS({\bf x},t)/\hbar},
\end{equation} 
where and $R$ and $S$ are real-valued, then the trajectory 
of the particle is determined by the relation
\begin{equation}
{\bf p}=\nabla S.
\label{basicbohm}  
\end{equation}
The {\em Schr\"{o}dinger current} associated with $\psi$ is given by
\begin{equation}
\label{schrodingercurrent}
{\bf j} = \frac{1}{m} \rho {\bf p}
\end{equation}
where $\rho = \psi^* \psi = R^2$.
Comprehensive discussions of this causal interpretation
of quantum mechanics can be found in \cite{BoHi} and \cite{Ho93}.

However, as Holland \cite{Ho99} has pointed out, 
Eqs. \eqref{basicbohm} and \eqref{schrodingercurrent}
are relevant only to spin-0 particles.  For particles with spin,
these equations are inconsistent if the theory
is to be ultimately embedded in a relativistic theory.
The condition of Lorentz covariance on the law of
motion implies that the momentum of a particle with spin {\bf s} must be
given by \cite{Ho99}
\begin{equation}\label{withterm}
{\bf p}=\nabla S + \nabla \log\rho \times {\bf s}.
\end{equation}
In a number of earlier works, including \cite{GuHe,He75,He79},
the current vector associated with Eq. \eqref{withterm},
\begin{equation}
\label{paulicurrent}
{\bf j} = \frac{1}{m} \rho \nabla S + \frac{1}{m} \nabla \rho \times {\bf s},
\end{equation}
was referred to as the {\em Pauli current},
the nonrelativistic limit of
the {\em Dirac current}, as opposed to Eq. \eqref{schrodingercurrent},
the nonrelativistic limit of the {\em Gordon current}.
(In these papers, it was claimed that consistency with Dirac theory
requires that Schr\"odinger theory be regarded as describing
an electron in an eigenstate of spin.)  
The spin-dependent term was also discussed in \cite{BoHi}
but only in the context of the Pauli equation and not the
Schr\"{o}dinger equation.

The momentum defined in Eq. \eqref{basicbohm} predicts that electrons in an
eigenstate are stationary since $\nabla S = 0$.  This is a
counterintuitive result of Bohm's original theory, but one which no
longer persists when the extra term in \eqref{withterm} is taken into
account.  For example, consider an electron in the
the $1s$ ground eigenstate of hydrogen,
\begin{equation}
\label{psi_100}
\psi_{100} = \frac{1}{\sqrt{\pi a^3}} e^{-r/a},
\end{equation}
where $a = \hbar^2/(m e^2)$ is the Bohr radius.
Also assume that the electron is in a definite spin eigenstate:
Without loss of generality, let its spin vector be given by
${\bf s} = \frac{\hbar}{2}{\bf k}$.   (We shall justify this assumption
below.)
Holland \cite{Ho93} showed that the
extra term $\nabla \log\rho \times {\bf s}$ implies that the 
polar coordinates $r$ and $\theta$ are constant, with angle $\phi$
evolving in time as follows:
\begin{equation}
\label{phi_100}
\frac{d\phi}{dt}= \frac{\hbar}{mar}.
\end{equation}
All points on a sphere of radius $r$ orbit
the $z$-axis at the same rate.
If $r=a$, the angular frequency is on the order of $10^{16} ~ \text{s}^{-1}$,
cf. Eq. \eqref{omega}.

In Section 2, we examine the effects of
the extra term in \eqref{withterm} for an electron in several
eigenstates of the hydrogen atom.
Because the momentum equation \eqref{withterm} now involves
spin, a complete description of the electron in the atom --
provided by an appropriate wavefunction -- will
have to involve both spatial as well as spin information.
Let us denote the complete wavefunction of the electron by
$\Psi({\bf x},{\bf s},t)$, where ${\bf s}$ denotes appropriate spin
coordinates.  Since the hamiltonian describing the evolution of
$\Psi$ is the simple spin-independent hydrogen atom hamiltonian $\hat{H}_0$, 
we may write $\Psi$ as the tensor product $\psi({\bf x},t)\zeta({\bf s})$,
where $\psi({\bf x},t)$ is a solution of the time-dependent Schr\"{o}dinger
equation,
\begin{equation}
\label{tdse}
i \hbar \frac{\partial \psi}{\partial t} = \hat{H}_0 \psi ,
\end{equation}
and $\zeta({\bf s})$ is an eigenfunction of the commuting spin operators
$\hat{S}^2$ and $\hat{S}_z$, with $\hat{S}^2\zeta = \frac{3}{4}\hbar \zeta$
and $\hat{S}_z \zeta = \frac{1}{2}\hbar \zeta$.  Thus $\zeta$ defines the
``alpha'' or ``spin up'' state to which 
corresponds the spin vector ${\bf s} = \frac{\hbar}{2}{\bf k}$.
As such, the remainder of our discussion can focus
on the evolution of the spatial portion of the wavefunction
$\psi({\bf x},t)$.  In the parlance of ``state preparation,''
one can view this construction as preparing the electon with 
constant spin vector ${\bf s}$
and initial wavefunction $\psi({\bf x},0)$.  
Given an initial position ${\bf x}_0$ of the electron, Eq. \eqref{withterm} 
will then determine its initial momentum ${\bf p}_0$, from which its
causal trajectory then evolves.

In Section 3, we examine the trajectories of an electron,
as dictated by Eq. \eqref{withterm},
where $\psi$ is a linear combination
of $1s$ and $2p_0$ eigenstates evolving in time under the hydrogen
atom hamiltonian, cf. Eq. \eqref{tdse}.
Once again because of the spin-independence of $\hat{H}_0$,
we may assume that the spin vector of the elctron is constant, i.e.,
${\bf s} = \frac{\hbar}{2}{\bf k}$, and thereby focus on the
time evolution of the spatial wavefunction $\psi({\bf x},t)$.
This simple model was chosen to determine the major qualitative
features of Bohmian trajectories for a
$1s$-$2p$ transition induced by an oscillating electric field.
We have analyzed trajectories corresponding to the
time-dependent wavefunction associated with the transition
hamiltonian and shall report the results elsewhere.

\section{Spin-dependent trajectories of electrons in hydrogen eigenstates}

\subsection{Qualitative features}
Following the discussion at the end of the previous section,
we consider an electron with spin vector ${\bf s}$ 
that begins in a hydrogenic eigenstate, i.e.,
$\psi({\bf x},0) = \psi_{nlm}({\bf x})$.  The time evolution of the
spatial wavefunction is simply
\begin{equation}
\label{evol}
\psi({\bf x},t) = \psi_{nlm}({\bf x})e^{-i E_n t / \hbar}.
\end{equation} 
Comparing Eqs. \eqref{evol} and \eqref{bohmpsi}, we see that
for a real eigenstate, $\nabla S = 0$ so that the momentum in Eq. \eqref{withterm} is given by
\begin{equation}
\label{justterm}
{\bf p}=\nabla \log\rho \times {\bf s} ~ .
\end{equation} 
Some simple qualitative information about the electron trajectories
is readily found from this equation.
First, the vector $\nabla \log\rho$ points in the direction of the steepest
increase in $\log\rho$, hence in 
$\rho=|\psi|^2$.
Because of the cross product, the momentum vector
{\bf p} is perpendicular to this direction. In other
words, the trajectories of the electron lie
on level surfaces of $|\psi|^2$. 
However, {\bf p} is also
perpendicular to the direction of the spin, assumed to lie along the
$z$-axis in this discussion.
This implies that {\it z} is constant for these Bohm trajectories.

From the above analysis,
the {\it shape} of the Bohm trajectories may be found by
computing level surfaces of $|\psi|^2$ -- or simply $\psi$ for
real-valued eigenfunctions --
and then finding the intersections
of these surfaces with planes of constant $z$.
An electron in the $1s$ state of hydrogen, cf. Eq. \eqref{psi_100}, 
must therefore execute a circular orbit about the $z$-axis.
However, the angular velocity of this orbit, cf. Eq. \eqref{phi_100}, 
cannot be determined from this analysis.

For the $2s$ case, with quantum numbers $(n,l,m)=(2,0,0)$, the
wavefunction is given by
\begin{equation}
\label{psi_200}
\psi_{200}   = \frac{1}{\sqrt{32 \pi a^5}}(1-\frac{r}{2a})e^{-r/2a} .
\end{equation}
The level surfaces of $\psi_{200}$ are spheres whose intersection
with planes of constant $z$ are circles (with constant $\theta$ values).
The electron again travels about the $z$-axis in a circular orbit.

In the $2p_0$ case, $(n,l,m)=(2,1,0)$,
the wavefunction is given by
\begin{equation}
\label{psi_210}
\psi_{210}= \frac{1}{\sqrt{32\pi a^5}} re^{-r/2a}\cos\theta .
\end{equation}
The condition that both $\psi$ and $z$ be constant
is satisfied only if both $r$ and $\theta$ are constant, once again yielding
circular orbits about the $z$-axis.  

In the other $2p$ cases, $(n,l,m)=(2,1,\pm 1)$,
the wavefunctions are given by
\begin{equation}
\label{psi_21pm}
\psi_{21(\pm 1)} = \mp \frac{1}{\sqrt{32 \pi a^5}}
r e^{-r/2a}\sin{\theta}e^{\pm i \phi}. 
\end{equation}
The condition $\rho = | \psi |^2 = \text{constant}$ yields a
relation $r = r(\theta)$ defined implicitly by
\begin{equation} 
\sin \theta = \frac{K}{r}e^{r/2a} , 
\end{equation}
where $K$ is a constant.
Since $z=r\cos\theta$ is also constant, it follows that
both $r$ and $\theta$ are constants of motion so that
there is circular motion about the $z$-axis.

\subsection{More detailed dynamical descriptions of the trajectories}

For more complicated hydrogen eigenstates (see below), it may not be
as simple to find closed form expressions for level sets of $|\psi|^2$ so
that the method of qualitative analysis outlined above 
may be difficult if not impossible.
As well, it is desirable to extract quantitative information
such as the angular velocity $d \phi / dt$ of the
circular orbits deduced earlier.
We therefore analyze the differential equations of motion defined by
Eq. \eqref{justterm}.  

The gradient term from Eq. \eqref{justterm} is given by
\begin{equation}
\label{grad_cmplx}
\nabla \log\rho = \nabla \log\psi^*\psi =
2 \text{Re} \left ( \frac{(\nabla\psi)\psi^*}{\psi\psi^*} \right ).
\end{equation}
For real wavefunctions, this simplifies to
\begin{equation}
\label{grad_real}
\nabla \log \rho = 2\frac{\nabla\psi}{\psi}. 
\end{equation}
In spherical polar coordinates $(r,\theta,\phi)$, 
the spin vector ${\bf s}=\frac{\hbar}{2}{\bf k}$
is given by 
${\bf s}=\frac{\hbar}{2}(\cos\theta\hat{r}-\sin\theta\hat{\theta})$.
It is convenient to compute the cross product in 
the (right-handed) spherical polar coordinate system:
\begin{equation}
 \mathbf{A}\times\mathbf{B}=
\begin{vmatrix} \hat{\theta} & \hat{\phi}& \hat{r}\\
                                A_{\theta} & A_{\phi} & A_{r} \\
                                B_{\theta} & B_{\phi} & B_{r}
\end{vmatrix}.
\end{equation}

For the $1s$ ground state defined in Eq. \eqref{psi_100}, 
$\nabla \psi = -\frac{1}{a}\psi \hat r$.  Thus
\begin{equation}
{\bf p} = \frac{\hbar}{a}\sin \theta \hat{\phi}.
\end{equation}
Since $p_r = p_\theta = 0$, it follows that $r$ and $\theta$ are constant,
implying that $z$ is constant, i.e., circular orbits about the $z$-axis.
Holland's result in Eq. \eqref{phi_100} follows.

For the $2s$ wavefunction defined in Eq. \eqref{psi_200},
\begin{equation}
\frac{\nabla\psi_{200}}{\psi_{200}}=-\frac{1}{2a}
\left[\frac{1}{1-\frac{r}{2a}}+1\right]\hat{r},
\end{equation}
so that
\begin{equation}
{\bf p}=-\frac{\hbar}{2a}\sin\theta
\left[\frac{1}{1-\frac{r}{2a}}+1\right]\hat{\phi}.
\end{equation}
Once again $r$ and $\theta$ are constant.
From the relation $d\phi/dt=p_{\phi}/mr\sin\theta$, we have 
\begin{equation}
\label{phi_200}
\frac{d\phi}{dt}=-\frac{\hbar}{2mar}
\left[\frac{1}{\frac{r}{2a}-1}-1\right].
\end{equation}
Note that the pole at $r=2a$ coincides with the zero of the $2s$ 
wavefunction, implying that the 
probability of finding the electron at $r=2a$ is zero.
Also note that (i) $\dot{\phi} > 0$ for $0 < r < 2a$,
(ii) $\dot{\phi} < 0$ for $2a < r < 3a$, (iii) $\dot{\phi}=0$
for $r=3a$ and (iv) $\dot{\phi} > 0$ for $r > 3a$.
For $r=a$, the angular velocity $\dot{\phi}$ is equal to that of
the $1s$ ground state, cf. Eq. \eqref{phi_100}.

For the $2p_0$ state defined in Eq. \eqref{psi_210},
\begin{equation}
\frac{\nabla\psi_{210}}{\psi_{210}}=\frac{1}{r}(1-\frac{r}{2a})\hat{r}-
\frac{\sin\theta}{r\cos\theta}\hat{\theta}. 
\end{equation}
From Eq. \eqref{justterm}, 
\begin{equation}
{\bf p}=\frac{\hbar\sin\theta}{2a}\hat{\phi}, 
\end{equation}
implying that
\begin{equation}
\label{phi_210}
\frac{d\phi}{dt}=\frac{\hbar}{2mar}.
\end{equation}
This is one-half the angular velocity
for the $1s$ ground state.

For the $2p$ states with $m=\pm 1$, cf. Eq. \eqref{psi_21pm}, 
$\nabla S$ is not identically zero  
and so we must use the momentum
equation (4). We find that  
\begin{equation}
{\bf p} = \pm\frac{\hbar\sin\theta}{2a} \hat{\phi} \nonumber .
\end{equation}
 
The coordinates $r$ and $\theta$ are constant, implying
circular orbits about the $z$-axis.  The orbital angular
velocity is given by
 
\begin{equation}
\frac{d \phi}{dt} = \pm \frac{\hbar}{2mar} \nonumber .
\end{equation}
 
It will be instructive (for an analysis of Eq. (29) below) to
examine the Bohm trajectories resulting from Eq. (10), i.e.
ignoring the $\nabla S$ term.
We find that
 \begin{equation}
\label{phi_equation}
{\bf p}= - \frac{\hbar}{r}
\left[\sin\theta(1-\frac{r}{2a})+\frac{\cos^2\theta}{\sin\theta}\right]
\hat{\phi} .
\end{equation}
Once again, $r$ and $\theta$ are constant, implying 
circular orbits about the $z$-axis.  However the orbital
angular velocity depends upon $r$ and $\theta$ in a more complicated fashion:
\begin{equation}
\frac{d\phi}{dt}=\frac{\hbar}{mr^2}\left[\frac{r}{2a}-1-
\cot^2\theta\right].
\end{equation}
For $\theta = \pi/2$, i.e., the $xy$ plane, $\dot \phi = 0$ for
$r = 2a$.  This implies that there is a ring of equilibrium points
on the $xy$ plane at which the electron is stationary.  For all other
nonzero $r$ values on the $xy$ plane the electron revolves about the $z$-axis:
$\dot \phi < 0$ for $0 < r < 2a$ and $\dot \phi > 0$ for $r > 2a$.
A ring of stationary points exists on every plane parallel to
the $xy$ plane:  For any fixed $\theta_0 \in (0, \pi/2)$,
the ring is determined by the relation $r = 2a \csc^2 \theta_0$.
The radius of this ring (distance from the $z$-axis) is 
$r \sin \theta_0 = 2a / \sin \theta_0$.
The set of all points at which
the electron is stationary defines a surface that is generated
by revolving the curves
\begin{equation}
\label{curves}
z = \pm x \left [ \left ( \frac{x}{2a} \right )^2 -1 \right ]^{1/2}, ~~~~
x \geq 2a,
\end{equation}
about the $z$-axis.
In the region between this surface and the $z$-axis, $\dot \phi < 0$;
on the other side of this surface, $\dot \phi > 0$.

In summary, for each of the hydrogen eigenstates studied
above, the Bohm trajectories are circular orbits about 
the $z$-axis, the assumed orientation of the electron spin vector ${\bf s}$.
Furthermore, the quantitative behaviour of the angular velocity
has been determined in all cases.

We now examine Bohm trajectories for the real hydrogen wavefunctions,
\cite{Le}
\begin{equation}
\begin{split}
\psi_{2p_x} &= N re^{-r/2a}\sin\theta\cos\phi  \\
\psi_{2p_y} &= N re^{-r/2a}\sin\theta\sin\phi ,
\end{split}
\end{equation}
where $N = 1/\sqrt{32 \pi a^5}$.  The probability distributions
associated with these wavefunctions are the familiar hydrogen orbitals
used in descriptions of organic chemical bonding.

The $\psi_{2p_x}$ and $\psi_{2p_y}$ wavefunctions are obtained by appropriate
linear combinations of the energetically degenerate
eigenfunctions $\psi_{21(\pm1)}$ 
of Eq. \eqref{psi_21pm}.  Therefore they are also eigenfunctions
of the hydrogen atom hamiltonian $\hat{H}_0$ with energy $E_2$.
From Eq. \eqref{evol}, it follows that $\nabla S = 0$.  The cross 
product of Eq. \eqref{justterm} has
components in all three variables, resulting in a system of three
coupled ordinary differential equations which must be integrated to
find the trajectories.

The system of ODEs associated with the 
$\psi_{2p_x}$ wavefunction is given by
\begin{equation}\label{eqs}
\begin{split}
\frac{dr}{dt} &= -\frac{\hbar}{mr}\tan\phi \\
\frac{d\theta}{dt} &= -\frac{\hbar}{mr^2}\cot\theta\tan\phi \\
\frac{d\phi}{dt} &= -\frac{\hbar}{mr^2}(1-\frac{r}{2a}+\cot^2\theta).
\end{split}
\end{equation}
Note that the $\phi$ DE is identical to Eq. \eqref{phi_equation}.
From the first two DEs, we have
\begin{equation}
\frac{dr}{d\theta} = r \tan \theta ,
\end{equation}
which is easily integrated to give $z = r \cos \theta = C$,
in agreement with our earlier analysis.
It follows that $r$ and $\theta$ are constant when
$\phi = 0$ or $\pi$.  If 
$\theta = \pi/2$ as well, then $\dot{\phi}=0$ for $r=2a$, implying the
existence of two equilibrium points at $(x,y,z)= (2a,0,0)$ and
$(-2a,0,0)$.  At these points, the electron is stationary.
In fact, these are two particular cases of an
infinity of equilibrium points that are given by 
the conditions $\phi=0,\pi$ ($xz$ plane) and
\begin{equation}
1-\frac{r}{2a}+\cot^2\theta=0.
\end{equation}
By virtue of the above relation and Eq. \eqref{phi_equation},
the equilibrium points of this system
lie on the two curves defined by Eq. \eqref{curves} in the $xz$ plane
as well as their reflections about the $z$-axis.  Each of
these points corresponds to the points of highest and
lowest ``elevation'' (from the horizontal $xy$ plane) 
of the familiar dumb-belled level
surfaces $\rho=\psi^2=C$ of the orbital.
At all of these points, the electron is stationary.

The system of ODEs in \eqref{eqs} may be integrated numerically.
However, it is useful to 
introduce the dimensionless variables 
$\xi = r/a$ and $\tau=\omega_0 t$, where 
\begin{equation}
\label{omega}
\omega_0 = \frac{E_2 - E_1}{\hbar} = \frac{3 \hbar}{8ma^2}
\end{equation}
is the angular frequency
associated with the $n=1$ to $n=2$ transition in the hydrogen atom
($\omega_0 \approx 1.549\times10^{16}~ \text{s}^{-1}$). The scaled equations become
\begin{equation}
\begin{split}
\frac{d\xi}{d\tau} &=
-\frac{8}{3\xi}\tan\phi\\
\frac{d\theta}{d\tau} &=
-\frac{8}{3\xi^2}\cot\theta\tan\phi\\
\frac{d\phi}{d\tau} &= -\frac{8}{3\xi^2}
(1-\frac{\xi}{2}+\cot^2\theta).
\end{split}
\end{equation}
Figure 1 shows the numerically integrated trajectories for 
several initial conditions in the $xz$ plane.
Note that there is a good qualitative agreement between these
trajectories and the orbital shapes of the $2p_x$ state as
depicted in textbook contour plots.
(No orbits cross the $yz$ plane since it is a nodal surface.)
The numerical results confirm that motion is periodic.  Angular frequency 
values are observed to be of the order of $\omega_0$.
These periodic orbits are stable in the sense of Lyapunov.

\begin{figure*}
\begin{center}
\epsfig{file=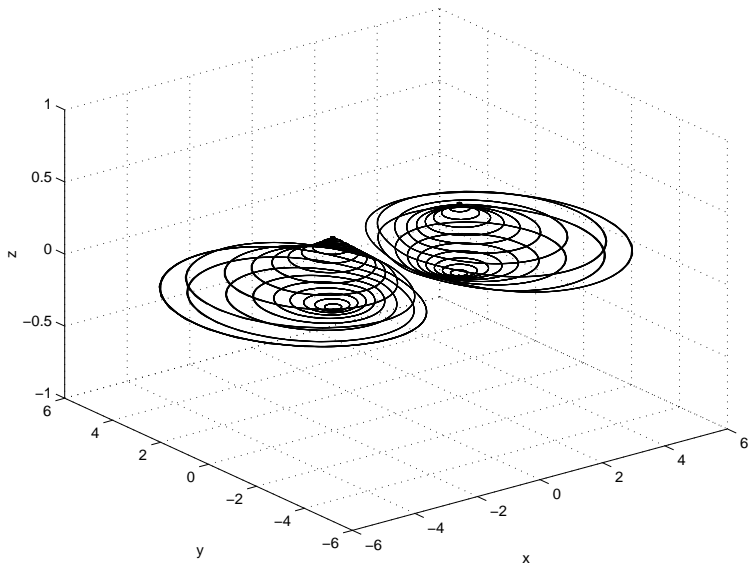,width=10cm}
\vspace{0.5cm}
\end{center}
\caption{Spin-dependent Bohm trajectories for the $2p_x$ hydrogen eigenstate}
\end{figure*}

The nondimensional and numerical analysis 
of the $2p_y$ case proceeds in a similar fashion.
The resulting system of ODEs represents a rotation of 
the system in Eq. \eqref{eqs}
by an angle of $\pi/2$ in $\phi$.

To summarize this section,
we have shown that the spin-dependent Bohm trajectories
for the ground and first excited states of the hydrogen atom
are stable periodic orbits.
There are no exceptional orbits that deviate from this regularity.
For some of the $2p$ states, these orbits include
families of stationary points that have zero 
Lebesgue measure in ${\bf R}^3$.
These results are intuitively more acceptable
than the original Bohmian result that {\it all} trajectories
associated with a given eigenstate are stationary.
\section{Trajectories associated with a linear 
superposition of hydrogenic eigenfunctions}
We now examine the Bohm trajectories
of an electron with constant spin vector ${\bf s} = \frac{\hbar}{2}{\bf k}$
but with a spatial wavefunction $\psi({\bf x},t)$ that
begins as a linear combination of
$1s$ and $2p_0$ hydrogenic eigenfunctions,
\begin{equation}
\label{linearcombination}
\psi({\bf x},0) = c_1 \psi_{100}({\bf x}) + c_2 \psi_{210}({\bf x}),
\end{equation}
where $c_1^2+c_2^2=1$.
(The assumption that the electron is in a well-defined ``spin up''
eigenstate was justified at the end of Section 1.)

The time evolution of this spatial wavefunction, as dictated by the
time-dependent Schr\"{o}dinger equation \eqref{tdse}, will be given by
\begin{equation}
\label{tdse1s2p0}
\psi({\bf x},t) = c_1 \psi_{100}({\bf x})e^{-iE_1 t/\hbar} +
                  c_2 \psi_{210}({\bf x})e^{-iE_2 t/\hbar} .
\end{equation}
This linear combination was chosen in order to examine 
some of the qualitative features of the trajectories
associated with the $1s$-$2p_0$ transition
in hydrogen induced by an oscillating electric field
of the form $\hat{z} E_0 \cos \omega t$.
Indeed, many of the qualitative features of this problem
are captured by this model, with the exception of additional
oscillations along invariant surfaces due to the oscillating field.
We have studied the trajectories of this transition problem in detail
and shall report the results elsewhere.

The wavefunction in Eq. \eqref{tdse1s2p0} is {\em not} a linear combination
of energetically degenerate states.  As such $\nabla S$, the first
term of the momentum in Eq. \eqref{withterm}, is nonzero.
It can be computed as follows:
\begin{equation}
\nabla S = \frac{\hbar}{\psi^* \psi}
\text{Im} [ (\nabla \psi)\psi^* ] .
\end{equation}
The results are
\begin{eqnarray}
\label{hydro1}
p_r & = &
- \frac{\hbar}{\psi^* \psi} c_1c_2 N_1 N_2 e^{-3r/2a}\cos \theta 
\sin \omega_0 t \left ( 1 + \frac{r}{2a} \right ) , \nonumber \\
p_\theta &=&
\frac{\hbar}{\psi^* \psi} c_1c_2 N_1 N_2 e^{-3r/2a}\sin \theta 
\sin \omega_0 t , \\
p_\phi & = & 0 . \nonumber
\end{eqnarray}
Here, $N_1$ and $N_2$ denote, respectively, 
the normalization factors of the $1s$ and $2p$ states
and $\omega_0$ is defined in Eq. \eqref{omega}.
This is the momentum of the original Bohmian formulation
${\bf p} = \nabla S$.  The angle $\phi$ is constant, implying
that there is no orbital motion about the $z$-axis.

The second term in the momentum equation \eqref{withterm} contributes only to
the $\phi$-momentum:
\begin{equation}
\label{hydro2}
p_\phi = -\frac{\hbar}{\psi^* \psi} [ Y \cos \theta + X \sin \theta ],
\end{equation}
where
\begin{eqnarray}
X &=& 
- \frac{1}{a} c_1^2 N_1^2 e^{-2r/a} + c_2^2N_2^2 r 
\left ( 1 - \frac{r}{2a} \right ) e^{-r} \cos^2 \theta + \nonumber \\
  & & ~~~~~~ c_1c_2N_1N_2 \left ( 1 - \frac{3r}{2a} e^{-3r/2a} \right ) 
\cos \theta \cos \omega_0 t , \nonumber \\
Y &=& - c_2^2 N_2^2 re^{-r} \sin \theta \cos \theta 
- c_1c_2N_1N_2 e^{-3r/2a} \sin \theta \cos \omega_0 t .
\end{eqnarray}
(The terms $X$ and $Y$ are the $r$ and $\theta$ components, respectively,
of $\nabla \log \psi^*\psi$.)
Once again, the spin-dependent momentum term implies
orbital motion about the $z$-axis.

The $p_r$ and $p_\theta$ equations of
\eqref{hydro1} along with $p_\phi$ in \eqref{hydro2} yield differential
equations in the coordinates $r$, $\theta$ and $\phi$.
Once again, we rewrite these DEs in terms of the dimensionless variables 
$\xi = r/a$ and $\tau=\omega_0 t$.  The net result is the
following system:
\begin{eqnarray}
\label{modelproblem}
\frac{d \xi}{d \tau} & = &
- \frac{\sqrt{2}}{3D} c_1c_2 \left ( 1 + \frac{\xi}{2} \right ) 
e^{-3 \xi/2} \cos \theta \sin \tau , \nonumber \\
\frac{d \theta}{d \tau} & = &
 \frac{\sqrt{2}}{3D\xi} c_1c_2 
e^{-3 \xi/2} \sin \theta \sin \tau , \\
\frac{d \phi}{d \tau} & = &
 \frac{8}{3D\xi} 
\left [ c_1^2e^{-2\xi} + \frac{1}{64}c_2^2 \xi^2 e^{-\xi} \cos^2\theta +
\frac{3}{8\sqrt{2}}c_1c_2 \xi e^{-3\xi/2}\cos \theta \cos \tau \right ] , \nonumber
\end{eqnarray}
where
\begin{equation}
D = c_1^2e^{-2\xi} + \frac{1}{32}c_2^2 \xi^2 e^{-\xi} \cos^2\theta +
\frac{1}{2\sqrt{2}}c_1c_2 \xi e^{-3\xi/2}\cos \theta \cos \tau .
\end{equation}

We first note the following two special cases for the
system of DEs in \eqref{modelproblem} (the primes denote
differentiation with respect to $\tau$):
\begin{enumerate}
\item
$(c_1,c_2)=(1,0)$ ($1s$ state):  $\xi^\prime = \theta^\prime = 0$,
and $\phi^\prime = 8/(3\xi)$, in agreement with Eq. \eqref{phi_100}.
\item
$(c_1,c_2)=(0,1)$ ($2p_0$ state):  $\xi^\prime = \theta^\prime = 0$,
and $\phi^\prime = 4/(3\xi)$, in agreement with Eq. \eqref{phi_210}.
\end{enumerate}
In these cases, 
the trajectories are simple circular orbits about the
$z$-axis, as expected.

More generally, the DEs in $\xi$ and $\theta$ are decoupled from the
$\phi$ DE.  From the former two DEs, we have
\begin{equation}
\frac{d \xi}{d \theta} = - \xi \left ( 1 + \frac{\xi}{2} \right ) \cot \theta .
\end{equation}
This separable DE is easily solved to give
\begin{equation}
\label{hyperbola}
\xi = \frac{2}{A \sin \theta - 1}, ~~~~~
A = \frac{2 + \xi_0}{\xi_0 \sin \theta_0} > 1 ,
\end{equation}
where $\xi_0 = \xi(0)$ and $\theta_0 = \theta(0)$.
In a plane with $z = \xi \cos \theta$ defining the vertical axis
(recall that $0 \leq \theta \leq \pi/2$),
the relation $\xi(\theta)$ defines a family of hyperbolae.
The asymptotes of the hyperbola in Eq. \eqref{hyperbola}
are given by the rays $\theta = \theta_1 = 
\mbox{Sin}^{-1} [\xi_0 \sin \theta_0/(\xi_0 + 2)]$ and 
$\theta = \theta_2 = \pi - \theta_1$.
If we choose $\theta_0 = \pi/2$ ($x$-axis), then
$\xi_0 \to 0^+$, $\theta_1 \to 0^+$, i.e., the hyperbolae
flatten as they approach the $z$-axis.

Since the $\xi$-$\theta$ DEs are decoupled from the $\phi$
equation in \eqref{modelproblem}, the electron will remain on the 3D surface
obtained by rotating the appropriate $\xi(\theta)$ hyperbola
in Eq. \eqref{hyperbola} about the $z$-axis.
In other words, the hyperboloid of revolution is an
invariant set in ${\bf R}^3$ for the electronic trajectory.
Plots of three sample trajectories for the case 
$c_1=c_2=\frac{1}{\sqrt{2}}$ are shown
in Figure 2.  The time interval $0 \leq \tau \leq 50$
was chosen so that the oscillatory nature of the solutions
could be seen.   

In Figure 3, the time interval of these
solutions has been extended to $0 \leq \tau \leq 1000$.
The invariant surfaces associated with these trajectories 
can be discerned from the plots.

\begin{figure*}[h]
\begin{center}
\psfig{file=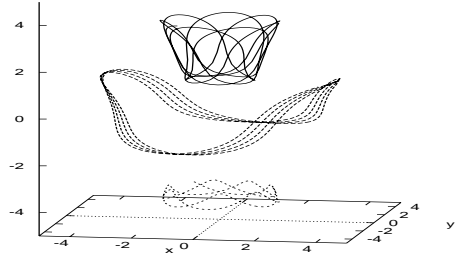,height=6cm,width=8cm}
\vspace{0.5cm}
\end{center}
\caption{Spin-dependent Bohm trajectories for the 
$1s$-$2p_0$ linear combination, $c_1 = c_2 = \frac{1}{\sqrt{2}}$, as computed
from Eq. \eqref{modelproblem}, for $0 \leq \tau \leq 50$.}
\end{figure*}

\begin{figure*}[h]
\begin{center}
\psfig{file=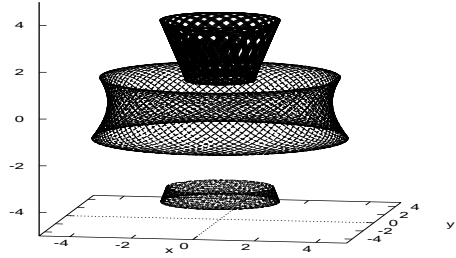,height=6cm,width=8cm}
\vspace{0.5cm}
\end{center}
\caption{The spin-dependent Bohm trajectories from Figure 2,
but computed for $0 \leq \tau \leq 1000$ in order to show
the invariant surfaces on which the trajectories lie.}
\end{figure*}

A more realistic simulation of the oscillating electric field
problem is accomplished if we set the coefficients $c_1$
and $c_2$ in Eq. \eqref{linearcombination} to be
\begin{equation}
\label{oscillatoryfield}
c_1 = \cos \omega t , ~~~ c_2 = \sin \omega t , ~~~ t \geq 0,
\end{equation}
so that the atom begins in the ground state and proceeds to oscillate
between it and the $2p_0$ excited state.
In this case, the electron trajectories, as determined by
by the system of ODEs in \eqref{modelproblem}, will
still be constrained to the invariant hyperboloid surfaces of
revolution.  However, there will be additional 
oscillatory components in these trajectories due to the 
periodic behaviour of the $c_i$ in Eq. \eqref{oscillatoryfield}.
A detailed examination of the possible qualitative behaviour
of solutions is beyond the scope of this letter.

\section*{Acknowledgements} We gratefully acknowledge that this
research has been supported by the Natural Sciences and Engineering
Research Council of Canada (NSERC) in the form of a Postgraduate
Scholarship (CC) and an Individual Research Grant
(ERV).

\end{document}